\documentclass{article}
\usepackage{amsmath}

\begin{document}

\begin{center}
\textbf{RELATIVELY MOVING SYSTEMS IN ``TRUE TRANSFORMATIONS RELATIVITY''}%
\bigskip

Tomislav Ivezi\'{c}\bigskip

\textit{Ru}%
\mbox
{\it{d}\hspace{-.15em}\rule[1.25ex]{.2em}{.04ex}\hspace{-.05em}}\textit{er Bo%
\v{s}kovi\'{c} Institute, P.O.B. 180, 10002 Zagreb, Croatia}

ivezic@irb.hr\bigskip
\end{center}

\noindent In this paper the physical systems consisting of relatively moving
subsystems are considered in the \textquotedblleft true transformations
relativity\textquotedblright . It is found in a manifestly covariant way
that there is a second-order electric field outside stationary
current-carrying conductor. It is also found that there are opposite charges
on opposite sides of a square loop with current and these charges are
invariant charges.\bigskip

\bigskip

\noindent \textit{Henceforth space by itself, and time by itself, are doomed
\newline
to fade away into mere shadows and only a kind of union of \newline
the two will preserve an independent reality.} - H. Minkowski\newline
\newline
\newline
\textbf{1. INTRODUCTION}\newline
\newline
In the recent paper [1] I have shown that due to the fundamental difference
between the true transformations (TT) and the apparent transformations (AT)
(see [1] and [2]) one can speak about two forms of relativity: the
\textquotedblleft TT relativity\textquotedblright\ and the \textquotedblleft
AT relativity.\textquotedblright\ The \textquotedblleft TT
relativity,\textquotedblright\ which is a covariant formulation of
relativity, is based on the TT of physical quantities as 4-dimensional (4D)
spacetime tensors, i.e., on the covariant definition of the spacetime
length, and the covariant electrodynamics with 4-vectors $E^{\alpha }$ and $%
B^{\alpha },$ see [1] and [3]. This formulation of electrodynamics is
equivalent to the usual covariant electrodynamics with the electromagnetic
field tensor $F^{\alpha \beta }$, as shown in [1]. \textit{The TT are the
transformations of 4D spacetime tensors referring to the same quantity (in
4D spacetime) considered in different inertial frames of reference (IFRs),
or in different coordinatizations of some IFR.} The TT do conform with the
special relativity as the theory of 4D spacetime with pseudo-Euclidean
geometry, i.e., they leave the interval $ds$ and thus the geometry of
spacetime unchanged. An example of the TT are the Lorentz transformations
(LT) of 4D tensor quantities. The \textquotedblleft AT
relativity\textquotedblright\ is the conventional special relativity based
on Einstein's relativity of simultaneity and on the synchronous definition
of the spatial length, i.e., on the AT of the spatial length (the Lorentz
contraction, see [1,2,3,4]) and the time distance (the conventional
dilatation of time), and, as shown in [1] (see also [3]), on the AT of the
electric and magnetic three-vectors (3-vectors) $\mathbf{E}$ and $\mathbf{B}$%
\/\ (the conventional transformations of $\mathbf{E}$ and $\mathbf{B}$).
\textit{The AT are not the transformations of 4D spacetime tensors and they
do not refer to the same quantity (in 4D spacetime), but, e.g., they refer
to the same measurement in different IFRs.}

In this paper we investigate physical systems consisting of relatively
moving subsystems, as it is a current-carrying conductor (CCC), using a
covariant formulation of physical quantities and physical phenomena, i.e.,
the \textquotedblleft TT relativity\textquotedblright . First we examine the
covariant definition of length when defined in geometrical terms and in
different coordinatizations of an IFR. We also report an expression for the
Lorentz transformations, which is independent of the chosen synchronization,
i.e., coordinatization of an IFR. Further, the AT of the spatial length -
the Lorentz contraction - is examined in detail. Then the covariant
definition of length in Einstein's coordinatization is applied to the
consideration of the well-known \textquotedblleft
relativistic\textquotedblright\ paradox \textquotedblleft Car and garage
paradox.\textquotedblright\ It is found that in the \textquotedblleft TT
relativity\textquotedblright\ and, \textit{if one wants to retain the
connection with the prerelativistic physics in which one deals with the
\textquotedblleft spatial length\textquotedblright }, then only the rest
length (volume) of the object is well defined quantity.

From this result and the covariant definition of charge we also find that in
the \textquotedblleft TT relativity\textquotedblright\ the charge density as
the three-dimensional (3D) quantity has definite physical meaning only for
charges at rest. In order to avoid from the beginning the misunderstanding
of the \textquotedblleft TT relativity\textquotedblright\ and of our choice
of the rest frame of the object, as the starting frame for the definitions
of 4D quantities, we emphasize that the \textquotedblleft TT
relativity\textquotedblright\ is covariant in the usual sense. \textit{In
the \textquotedblleft TT relativity\textquotedblright\ one can define 4D
physical quantities and investigate physical laws connecting such 4D
quantities in any IFR, not only in the rest frame of the object. The LT will
correctly connect the results of measurements of}\emph{\ }\textit{the same
4D quantity in two, arbitrary, relatively moving IFRs.} Thus, the
\textquotedblleft TT relativity\textquotedblright\ does not use a preferred
reference frame. Our choice of the rest frame of the object does not mean in
any way that this frame is a preferred IFR. The rest frame is, in fact, the
most convenient for the purpose of comparison with the prerelativistic
physics, in which one does not deal with 4D quantities but with
\textquotedblleft 3+1\textquotedblright\ quantities (the quantities defined
in \textquotedblleft 3+1\textquotedblright\ space and time), and with the
\textquotedblleft AT relativity,\textquotedblright\ in which one works in 4D
spacetime but with quantities, e.g., the spatial length, the time distance,
the 3-vectors \textbf{E }and\textbf{\ B, }etc\textbf{.,} that are not 4D
tensor quantities. Taking this into account we show that the current density
4-vector $j^{\mu }$ for a CCC in an arbitrary IFR is determined as the sum $%
j_{+}^{\mu }+j_{-}^{\mu }$, where the current density 4-vectors $j_{+}^{\mu }
$ and $j_{-}^{\mu }$ for positive and negative charges, respectively, have
to be found in their own rest frames, and then transformed by the Lorentz
transformation to the considered IFR. Then in Sec.3.2 we quote the covariant
Maxwell equations when written by the electromagnetic field tensor $%
F^{\alpha \beta }$ and by the 4-vectors $E^{\alpha }$ and $B^{\alpha },$
(both forms were already found in [1]), and also we report a new form - the
covariant Majorana form of Maxwell's equations. Then the 4-vectors $%
E^{\alpha }$ and $B^{\alpha }$ are determined for a CCC (instead of the
usual 3-vectors $\mathbf{E}$ and $\mathbf{B)}$ and it is obtained in such a
covariant way, i.e., in the \textquotedblleft TT
relativity,\textquotedblright\ that, for the observers at rest in the rest
frame of that CCC, there is a second-order electric field outside stationary
conductor with steady current. Such fields are already theoretically
predicted on different grounds in [5], see also [6]. In contrast to previous
works we also find in such a covariant manner that there are opposite
charges on opposite sides of a square loop with current and these charges
are Lorentz invariant charges. In the usual approach, i.e., in the
\textquotedblleft AT relativity,\textquotedblright\ it is found that there
is an electric moment $\mathbf{P}$\textbf{\ }for a moving loop with a
current. However we find that such loop, regarding the electric effects,
\emph{always}, i.e., for a stationary loop as well, behaves at long
distances as an electric dipole (4-vector).\newline
\newline
\newline
\textbf{2. COVARIANT AND SYNCHRONOUS DEFINITIONS OF \newline
LENGTH}\newline
\newline
As discussed in [1], (and [3]) according to the \textquotedblleft
modern\textquotedblright\ point of view the special relativity is the theory
of 4D spacetime with pseudo-Euclidean geometry. Quantities of physical
interest, both local and nonlocal, are represented in the special relativity
by spacetime tensors, i.e., as covariant quantities, and the laws of physics
are written in a manifestly covariant way as tensorial equations. The
geometry of the spacetime is generally defined by the invariant
infinitesimal spacetime distance $ds$ of two neighboring points, $%
ds^{2}=dx^{a}g_{ab}dx^{b}$. I adopt the following convention with regard to
indices. Repeated indices imply summation. Latin indices $a,b,c,d,...$ are
to be read according to the abstract index notation, see [7], Sec.2.4.. They
designate geometric objects in 4D spacetime. Thus $dx^{a,b}$ and $g_{ab}$,
and of course $ds,$ are defined independently of any coordinate system,
e.g., $g_{ab}$ is a second-rank covariant tensor (whose Riemann curvature
tensor $R_{bcd}^{a}$ is everywhere vanishing; the spacetime of special
relativity is a flat spacetime, and this definition includes not only the
IFRs but also the accelerated frames of reference). Greek indices run from 0
to 3, while latin indices $i,j,k,l,...$ run from 1 to 3, and they both
designate the components of some geometric object in some coordinate chart,
e.g., $x^{\mu }(x^{0},x^{i})$ and $x^{\prime \mu }(x^{\prime 0},x^{\prime i})
$ are two coordinate representations of the position 4-vector $x^{a}$ in two
different inertial coordinate systems $S$ and $S^{\prime },$ and $g_{\mu \nu
}$ is the $4\times 4$ matrix of components of $g_{ab}$ in some coordinate
chart.\newline
\newline
\newline
\textbf{2.1. The Spacetime or the TT Length}\newline
\newline
In general, in 4D spacetime of special relativity it is not possible to
separate the spatial and temporal parts of $ds$, or according to Minkowski's
words, quoted here as a motto, \textit{the spatial and temporal parts taken
separately loose their physical meaning}. Therefore, only the invariant
spacetime length (the Lorentz scalar) between two points (events) in 4D
spacetime does have definite physical meaning in the \textquotedblleft TT
relativity\textquotedblright\ and it is defined as

\begin{equation}
l=(l^{a}g_{ab}l^{b})^{1/2},  \label{covlen}
\end{equation}
where $l^{a}(l^{b})$ is the distance 4-vector between two events $A$ and $B$%
, $l^{a}=x_{B}^{a}-x_{A}^{a}$, $x_{A,B}^{a}$ are the position 4-vectors, and
$g_{ab}$ is the metric tensor.

Using different coordinatizations of a given reference frame, which can be
realized, for example, by means of different synchronizations, we find
different expressions, i.e., different representations of the spacetime
length $l,$ Eq.(\ref{covlen}). Obviously the coordinates $x^{\mu }$ of an
event, when written in some coordinatization of an IFR, do not have an
intrinsic meaning in 4D spacetime. However the spacetime length $l$ (\ref%
{covlen}) does have the same value for all relatively moving inertial
coordinate systems and it represents an intrinsic feature of the spacetime.

Different synchronizations are determined by the parameter $\varepsilon $ in
the relation $t_{2}=t_{1}+\varepsilon (t_{3}-t_{1})$, where $t_{1}$ and $%
t_{3}$ are the times of departure and arrival, respectively, of the light
signal, read by the clock at $A$, and $t_{2}$ is the time of reflection at $%
B $, read by the clock at $B$, that has to be synchronized with the clock at
$A $. Usually physicists prefer Einstein's synchronization convention with $%
\varepsilon =1/2$ in which the measured coordinate velocity of light (the
one-way speed of light) is constant and isotropic. A nice example of a
non-standard synchronization is \textquotedblleft
everyday\textquotedblright\ clock synchronization [8] in which $\varepsilon
=0$ and there is an absolute simultaneity; see also [9] for an absolute
simultaneity in the special relativity, and for the review on
synchronisation and test theories see the recent article [10]. As explained
in [8]: \textquotedblleft For if we turn on the radio and set our clock by
the standard announcement \textquotedblright ...at the sound of the last
tone, it will be 12 o'clock,\textquotedblright\ then we have synchronized
our clock with the studio clock in a manner that corresponds to taking $%
\varepsilon =0$ in $t_{2}=t_{1}+\varepsilon (t_{3}-t_{1})$.\textquotedblright

When Einstein's synchronization of distant clocks and cartesian space
coordinates $x_{e}^{i}$ are used in an IFR $S$ (this coordinatization will
be named Einstein's or \textquotedblleft e\textquotedblright\
coordinatization) then, e.g., the geometric object $g_{ab}$ is represented
by the $4\times 4$ matrix of components of $g_{ab}$ in that coordinate
chart, i.e., it is the Minkowski metric tensor $g_{\mu \nu e}=diag(-1,1,1,1)$%
, where \textquotedblleft e\textquotedblright\ stands for Einstein's
coordinatization. With such $g_{\mu \nu e}$ the space $x_{e}^{i}$ and time $%
t_{e}$ $(x_{e}^{0}\equiv ct_{e})$ components of $x_{e}^{\mu }$ do have their
usual meaning. Then $ds^{2}$ can be written with the separated spatial and
temporal parts, $ds^{2}=(dx_{e}^{i}dx_{ie})-(dx_{e}^{0})^{2}$, and the same
happens with the spacetime length $l$ (\ref{covlen})$,$ $%
l^{2}=(l_{e}^{i}l_{ie})-(l_{e}^{0})^{2}$. Such separation remains valid in
other inertial coordinate systems with the Minkowski metric tensor, and in $%
S^{\prime }$ one finds $l^{\prime 2}=(l_{e}^{\prime i}l_{ie}^{\prime
})-(l_{e}^{\prime 0})^{2}$, where $l_{e}^{\prime \mu }$ in $S^{\prime }$ is
connected with $l_{e}^{\mu }$ in $S$ by the LT.

In the usual form the LT connect two coordinate representations (in the
\textquotedblleft e\textquotedblright\ coordinatization) $x_{e}^{\mu }$, $%
x_{e}^{\prime \mu }$ of a given event. $x_{e}^{\mu }$, $x_{e}^{\prime \mu }$
refer to two relatively moving IFRs (with the Minkowski metric tensor) $S$
and $S^{\prime },$
\begin{equation*}
x_{e}^{\prime \mu }=L^{\mu }{}_{\nu ,e}x_{e}^{\nu
},\,\,\,L^{0}{}_{0,e}=\gamma _{e},L^{0}{}_{i,e}=L^{i}{}_{0,e}=-\gamma
_{e}V_{e}^{i}/c,L^{i}{}_{j,e}=\delta _{j}^{i}+(\gamma
_{e}-1)V_{e}^{i}V_{je}/V_{e}^{2},
\end{equation*}%
where $V_{e}^{i}=dx_{e}^{i}/dt_{e}$ are the components of the ordinary
velocity 3-vector, and $\gamma _{e}\equiv (1-V_{e}^{2}/c^{2})^{1/2}$. As
explained in [11], when such usual representations of pure Lorentz
transformations are applied to covariant expressions they destroy the
covariant form : \textquotedblleft because they employ three-vector
notation, because they treat the spatial and temporal components separately,
and because they are parametrized by the ordinary velocity three-vector $%
\mathbf{V}$.\textquotedblright\ In order to obtain a covariant expression
for $L^{\mu }{}_{\nu ,e}$ the ordinary velocity is replaced in [11] by the
proper velocity 4-vector $v_{e}^{\mu }\equiv dx_{e}^{\mu }/d\tau =(\gamma
_{e}c,\gamma _{e}v_{e}^{i})$, $d\tau \equiv dt_{e}/\gamma _{e}$ is the
scalar proper-time, the unit vector $n_{e}^{\mu }\equiv (1,0,0,0)$ along the
temporal axis is introduced, and $\delta _{j}^{i}$ is replaced with the
Minkowski metric tensor $g^{\mu }{}_{\nu e}$. This shows that the cartesian
space coordinates $x_{e}^{i}$ and Einstein's synchronization of distant
clocks are explicitly chosen in [11]. In such a way the covariant expression
for $L^{\mu }$ $_{\nu ,e}$ in the \textquotedblleft e\textquotedblright\
coordinatization is found in [11], Eq.(5),
\begin{equation*}
L^{\mu }{}_{\nu ,e}\equiv L^{\mu }{}_{\nu ,e}(v)=g^{\mu }{}_{\nu e}-\frac{%
2n_{e}^{\mu }v_{\nu e}}{c}+\frac{(n_{e}^{\mu }+v_{e}^{\mu }/c)(n_{\nu
e}+v_{\nu e}/c)}{1-n_{e}\cdot v_{e}/c}.
\end{equation*}%
Since we want to use the LT in different coordinatizations we generalize the
expression for $L^{\mu }{}_{\nu ,e}$ from [11] and find
\begin{equation}
L^{a}{}_{b}\equiv L^{a}{}_{b}(v)=g^{a}{}_{b}-\frac{2n^{a}v_{b}}{c}+\frac{%
(n^{a}+v^{a}/c)(n_{b}+v_{b}/c)}{1-n\cdot v/c}.  \label{fah}
\end{equation}%
Such form (\ref{fah}) of the LT can be applied to an arbitrary inertial
coordinate system in which the metric tensor can be different than the
Minkowski metric tensor, and thus the form of the covariant 4D Lorentz
transformations (\ref{fah}) is independent of the chosen synchronization,
i.e., coordinatization of reference frames. But we have to note that $n^{a}$
in (\ref{fah}) is a specific quantity. Namely it always has to be taken as
the unit vector along the temporal axis in the chosen IFR and the chosen
coordinatization. Nevertheless $L^{a}{}_{b}$ correctly transforms some 4D
tensor quantity from an IFR to another relatively moving IFR. For example,
when $L^{a}{}_{b}$ is applied to the position 4-vector $x^{a}$ one finds (in
the abstract index notation)
\begin{equation}
x^{\prime a}=x^{a}+\frac{\left[ n\cdot x-(2\gamma +1)v\cdot x/c\right]
n^{a}+(n\cdot x+v\cdot x/c)v^{a}/c}{1-n\cdot v/c}.  \label{fahn2}
\end{equation}

Let us examine the relations (\ref{fah}) and (\ref{fahn2}) in two different
coordinatizations. First in the \textquotedblleft e\textquotedblright\
coordinatization, in which the Minkowski metric tensor is used, $n^{a}$
becomes $n_{e}^{\mu }\equiv (1,0,0,0)$, $v_{e}^{\mu }=(\gamma _{e}c,\gamma
_{e}v_{e}^{i})$, and $\gamma _{e}=-n_{e}^{\mu }v_{\mu e}/c$, as in [11].
From the general relation $v^{a}v_{a}=-c^{2}$ one finds, in the
\textquotedblleft e\textquotedblright\ coordinatization, that $%
v_{e}^{0}=(c^{2}+v_{e}^{i}v_{ie})^{1/2},$ which shows that the expression
for $L^{\mu }{}_{\nu ,e}$ is parametrized essentially by the three spatial
components $v_{e}^{i}$ of the proper velocity 4-vector $v_{e}^{\mu }$. Then,
using the above expressions for $n_{e}^{\mu }$, $v_{e}^{\mu }$, and $\gamma
_{e}$ one finds from (\ref{fah}) and (\ref{fahn2}) the usual expressions for
pure LT, as in [11], i.e., the above mentioned $L^{\mu }{}_{\nu ,e}$ and $%
x_{e}^{\prime \mu }$, but with $v_{e}^{i}$ replacing the components of the
ordinary velocity 3-vector $\mathbf{V.}$ Also, we find the above mentioned
usual expressions in the \textquotedblleft e\textquotedblright\
coordinatization for $ds^{2}=ds_{e}^{2}=(dx_{e}^{i}dx_{ie})-(dx_{e}^{0})^{2}$
and $l^{2}=l_{e}^{2}=(l_{e}^{i}l_{ie})-(l_{e}^{0})^{2}$,with the separated
spatial and temporal parts.

In the similar way we use the relations (\ref{fah}) and (\ref{fahn2}) to
write the corresponding expressions in another coordinatization,
\textquotedblleft r\textquotedblright\ coordinatization, of an IFR, which is
found in [8], where \textquotedblleft everyday\textquotedblright\ or
\textquotedblleft radio\textquotedblright\ synchronization of distant clocks
is used. For simplicity we consider 2D spacetime as in [8]. Then the metric
tensor $g_{ab}$ becomes $g_{\mu \nu r}=\left(
\begin{array}{cc}
-1 & -1 \\
-1 & 0%
\end{array}%
\right) ,$ where \textquotedblleft r\textquotedblright\ stands for
\textquotedblleft radio\textquotedblright\ (it differs from that one in [8]
since the Minkowski tensors are different). The LT $L^{\mu }\,_{\nu ,r}$ in
the \textquotedblleft r\textquotedblright\ coordinatization can be easily
found from (\ref{fah}), from the known $g_{\mu \nu r}$, with $n_{r}^{\mu
}=(1,0)$ and $\gamma _{r}=-n_{r}^{\mu }v_{\mu r}/c=\gamma _{e}$. These
relations can be found as in [8], or by means of the matrix $T^{\mu }\,_{\nu
}$, which is given below. Thus the pure Lorentz transformation matrix $%
L^{a}{}_{b}$ (\ref{fah}) becomes in the \textquotedblleft
r\textquotedblright\ coordinatization
\begin{equation*}
L^{\mu }\,_{\nu ,r}=\left(
\begin{array}{ll}
K & 0 \\
-\beta _{r}/K & 1/K%
\end{array}%
\right) .
\end{equation*}%
Also we find the \textquotedblleft r\textquotedblright\ representation $%
x_{r}^{\prime \mu }$ of $x^{\prime a}$ (\ref{fahn2}),
\begin{equation*}
x_{r}^{\prime 0}=Kx_{r}^{0},\,\,x_{r}^{\prime 1}=(1/K)(-\beta
_{r}x_{r}^{0}+x_{r}^{1}),
\end{equation*}%
where $K=(1+2\beta _{r})^{1/2}$ and $\beta _{r}=dx_{r}^{1}/dx_{r}^{0}$ is
the velocity of the frame $S^{\prime }$ as measured by the frame $S.$
Further $ds^{2}=dx^{a}g_{ab}dx^{b}$ becomes in the \textquotedblleft
r\textquotedblright\ coordinatization $ds^{2}=ds_{r}^{2}=-\left[
(dx_{r}^{0})^{2}+2dx_{r}^{0}dx_{r}^{1}\right] $. We see that in the
\textquotedblleft r\textquotedblright\ coordinatization the spatial and
temporal parts of $ds^{2}$ are not separated, that is different than in the
coordinatization with the Minkowski metric tensor. The same holds for the
spacetime length $l,$ which is in the \textquotedblleft r\textquotedblright\
coordinatization determined as $l^{2}=l_{r}^{2}=-\left[
(l_{r}^{0})^{2}+2l_{r}^{0}l_{r}^{1}\right] $. Expressing $dx_{r}^{\mu }$, or
$l_{r}^{\mu }$, in terms of $dx_{e}^{\mu }$, or $l_{e}^{\mu }$ (the
transformation matrix between \textquotedblleft r\textquotedblright\ and
\textquotedblleft e\textquotedblright\ coordinatizations is
\begin{equation*}
T^{\mu }\,_{\nu }=\left(
\begin{array}{cc}
1 & -1 \\
0 & 1%
\end{array}%
\right) ,
\end{equation*}%
whence, e.g., $x_{r}^{0}=x_{e}^{0}-x_{e}^{1}$, $x_{r}^{1}=x_{e}^{1}$, and $%
\beta _{r}=\beta _{e}/(1-\beta _{e})$, see [11]) one finds that $%
ds_{r}^{2}=ds_{e}^{2}$, and also, $l_{r}^{2}=l_{e}^{2}$, as it must be.

The whole preceding discussion about the geometric quantities $x^{a}$, $%
l^{a} $, $ds$, $l$, .. and their different representations can be
illustrated in a way which better clarifies the difference between two sorts
of quantities. Again we consider the TT length (\ref{covlen}) (we use the
words - the TT length, the spacetime length, and the covariantly defined
length as synonyms) in two relatively moving IFRs $S$ and $S^{\prime }$ and
in two coordinatizations \textquotedblleft e\textquotedblright\ and
\textquotedblleft r\textquotedblright\ in these IFRs. Now, let the spacetime
be endowed with base vectors, the temporal and the spatial base vectors. The
bases $\left\{ e_{\mu }\right\} $, with the base vectors $\{e_{0},e_{1}\}$,
and $\left\{ r_{\mu }\right\} $, with the base vectors $\left\{
r_{0},r_{1}\right\} $, are associated with \textquotedblleft
e\textquotedblright\ and \textquotedblleft r\textquotedblright\
coordinatizations, respectively, of a given IFR. The temporal base vector $%
e_{0}$ is the unit vector directed along the world line of the clock at the
origin. The spatial base vector by definition connects \textit{simultaneous}
events, the event \textquotedblleft clock at rest at the origin reads 0
time\textquotedblright\ with the event \textquotedblleft clock at rest at
unit distance from the origin reads 0 time\textquotedblright , and thus it
is synchronization-dependent. The spatial base vector $e_{1}$ connects two
above mentioned simultaneous events when Einstein's synchronization ($%
\varepsilon =1/2$) of distant clocks is used. The temporal base vector $%
r_{0} $ is the same as $e_{0}$. The spatial base vector $r_{1}$ connects two
above mentioned simultaneous events when \textquotedblleft
everyday\textquotedblright\ clock synchronization ($\varepsilon =0$) of
distant clocks is used. All the spatial base vectors $r_{1}$, $r_{1}^{\prime
}$, .. are parallel and directed along an (observer-independent) light line.
Hence, two events that are everyday (\textquotedblleft r\textquotedblright )
simultaneous in $S$ are also \textquotedblleft r\textquotedblright\
simultaneous for all other IFRs. The connection between the bases $\left\{
e_{\mu }\right\} $ and $\left\{ r_{\mu }\right\} $ is\quad $r_{0}=e_{0}$,$%
\;r_{1}=e_{0}+e_{1}$, see [8]. Then the geometrical quantity, e.g., the
distance 4-vector $l_{AB}^{a}$ between two events $A$ and $B$, will be
represented by the vector in 2D spacetime, which have different
decompositions, representations, with respect to $\left\{ e_{\mu }\right\} $%
, $\left\{ e_{\mu }^{\prime }\right\} $ and $\left\{ r_{\mu }\right\} $, $%
\left\{ r_{\mu }^{\prime }\right\} $ bases. \textit{Note that in the
\textquotedblleft TT relativity\textquotedblright\ the same distance
4-vector }$l_{AB}^{a}$ \textit{is considered (measured) in different
relatively moving IFRs and in different coordinatizations of these IFRs.}

In order to retain the connection with the prerelativistic physics and to
facilitate the comparison with the \textquotedblleft AT
relativity\textquotedblright\ we consider a particular choice for the
4-vector $l_{AB}^{a}$ (in the usual \textquotedblleft 3+1\textquotedblright\
picture it corresponds to an object, a rod, that is at rest in an IFR $S$
and situated along the common $x_{e}^{1},x_{e}^{\prime 1}-$ axes). In the
\textquotedblleft e\textquotedblright\ coordinatization the position
4-vectors of the events $A$, $x_{A}^{a}$, and $B$, $x_{B}^{a}$, in $S$ are
decomposed with respect to $\left\{ e_{\mu }\right\} $ base as $%
x_{A}^{a}=x_{Ae}^{0}e_{0}+x_{Ae}^{1}e_{1}=0e_{0}+0e_{1}$, and $%
x_{B}^{a}=x_{Be}^{0}e_{0}+x_{Be}^{1}e_{1}=0e_{0}+l_{0}e_{1}$, and the
distance 4-vector $l_{AB}^{a}=x_{B}^{a}-x_{A}^{a}$ is decomposed as
\begin{equation*}
l_{AB}^{a}=l_{e}^{0}e_{0}+l_{e}^{1}e_{1}=0e_{0}+l_{0}e_{1}.
\end{equation*}%
Thus in $S$ the position 4-vectors $x_{A,B}^{a}$ are determined
simultaneously, $x_{Be}^{0}-x_{Ae}^{0}=l_{e}^{0}=0$, i.e., the temporal part
of $l_{AB}^{a}$ is zero. The spacetime length $l$ is written in the $\left\{
e_{\mu }\right\} $ base as
\begin{equation*}
l=l_{e}=(l_{e}^{\mu }l_{\mu e})^{1/2}=(l_{e}^{i}l_{ie})^{1/2}=l_{0},
\end{equation*}%
as in the prerelativistic physics; it is in that case a measure of the
spatial distance, i.e., of the rest spatial length of the rod. The observers
in all other IFRs will look at the same events but associating with them
different coordinates; it is the essence of the covariant description. They
all obtain the same value $l$ for the spacetime length. It has to be pointed
out that \textit{in the \textquotedblleft TT relativity\textquotedblright\
it is not necessary to start in this example with the rest frame of the
object and to choose the events }$A$\textit{\ and }$B$\textit{\ to be
simultaneous in that frame. The whole consideration can be done in the same
covariant manner for other choices of IFRs and of the events }$A$\textit{\
and }$B$\textit{\ in the chosen IFR}. For any starting choice the covariant
LT (\ref{fah}) will correctly connect the results of measurements of \textit{%
the same 4D quantity} in two relatively moving IFRs. The rest frame of the
object and the simultaneity of the events $A$ and $B$ in it are chosen only
to have the connection with the prerelativistic physics, which deals with
\textquotedblleft 3+1\textquotedblright\ quantities and not with 4D
quantities.

Let us then consider the same 4-vector $l_{AB}^{a}$ in $S^{\prime }$, (where
in \textquotedblleft 3+1\textquotedblright\ picture the rod is moving). The
position 4-vectors $x_{A}^{a}$ and $x_{B}^{a}$ of the events $A$ and $B$
respectively are decomposed with respect to $\left\{ e_{\mu }^{\prime
}\right\} $ base as $x_{A}^{a}=x_{Ae}^{\prime 0}e_{0}^{\prime
}+x_{Ae}^{\prime 1}e_{1}^{\prime }=0e_{0}^{\prime }+0e_{1}^{\prime }$, and $%
x_{B}^{a}=x_{Be}^{\prime 0}e_{0}^{\prime }+x_{Be}^{\prime 1}e_{1}^{\prime
}=-\beta _{e}\gamma _{e}l_{0}e_{0}^{\prime }+\gamma _{e}l_{0}e_{1}^{\prime }$%
, and the distance 4-vector is decomposed as
\begin{equation*}
l_{AB}^{a}=x_{B}^{a}-x_{A}^{a}=l_{e}^{\prime 0}e_{0}^{\prime }+l_{e}^{\prime
1}e_{1}^{\prime }=-\beta _{e}\gamma _{e}l_{0}e_{0}^{\prime }+\gamma
_{e}l_{0}e_{1}^{\prime }.
\end{equation*}%
Note that in the \textquotedblleft e\textquotedblright\ coordinatization,
commonly used in the \textquotedblleft AT relativity,\textquotedblright\
there is a dilatation of the spatial part $l_{e}^{\prime 1}=\gamma _{e}l_{0}$
with respect to $l_{e}^{1}=l_{0}$ and not the Lorentz contraction as
predicted in the \textquotedblleft AT relativity.\textquotedblright\ Hovewer
it is clear from the above discussion that comparison of only spatial parts
of the components of the distance 4-vector $l_{AB}^{a}$ in $S$ and $%
S^{\prime }$ is physically meaningless in the \textquotedblleft TT
relativity.\textquotedblright\ All components of the distance 4-vector $%
l_{AB}^{a}$ are transformed by the LT from $S$ to $S^{\prime }$. $l_{e}^{\mu
}$ and $l_{e}^{\prime \mu }$ are different representations of the same
physical quantity $l_{AB}^{a}$ measured in two relatively moving IFRs $S$
and $S^{\prime }$. The invariant spacetime length of that object in $%
S^{\prime }$ is
\begin{equation*}
l=l_{e}^{\prime }=(l_{e}^{^{\prime }\mu }l_{\mu e}^{\prime })^{1/2}=l_{0}.
\end{equation*}%
Note that if $l_{e}^{0}=0$ then $l_{e}^{\prime \mu }$ in any other IFR $%
S^{\prime }$ will contain the time component $l_{e}^{\prime 0}\neq 0$. We
conclude from the above discussion that if one wants in the
\textquotedblleft TT relativity\textquotedblright\ to compare in a
physically meaningful sense the \textquotedblleft lengths\textquotedblright\
of two different objects than it is possible only by comparing their
invariant spacetime lengths.

In the \textquotedblleft r\textquotedblright\ coordinatization the position
4-vectors of the events $A$ and $B$, $x_{A}^{a}$ and $x_{B}^{a}$, in $S$ are
decomposed with respect to $\left\{ r_{\mu }\right\} $ base as $%
x_{A}^{a}=x_{Ar}^{0}r_{0}+x_{Ar}^{1}r_{1}=0r_{0}+0r_{1}$, and $%
x_{B}^{a}=x_{Br}^{0}r_{0}+x_{Br}^{1}r_{1}=-l_{0}r_{0}+l_{0}r_{1}$, and the
distance 4-vector $l_{AB}^{a}=x_{B}^{a}-x_{A}^{a}$ is decomposed as
\begin{equation*}
l_{AB}^{a}=l_{r}^{0}r_{0}+l_{r}^{1}r_{1}=-l_{0}r_{0}+l_{0}r_{1},
\end{equation*}%
and the TT length $l$ is
\begin{equation*}
l=l_{r}=(l_{r}^{\mu }l_{\mu r})^{1/2}=l_{e}=l_{0}
\end{equation*}%
as it must be.

In $S^{\prime }$ and in the $\left\{ r_{\mu }^{\prime }\right\} $ base the
position 4-vectors of the events $A$ and $B$ are $x_{A}^{a}=0r_{0}^{\prime
}+0r_{1}^{\prime }$ and $x_{B}^{a}=x_{Br}^{\prime 0}r_{0}^{\prime
}+x_{Br}^{\prime 1}r_{1}^{\prime }=-Kl_{0}r_{0}^{\prime }+(1+\beta
_{r})(1/K)l_{0}r_{1}^{\prime }$, and the components $l_{r}^{\prime \mu }$ of
the distance 4-vector $l_{AB}^{a}$ are equal to the components $%
x_{Br}^{\prime \mu }$, i.e., $l_{r}^{\prime \mu }=x_{Br}^{\prime \mu }$.
Thus $l_{AB}^{a}$ is decomposed as
\begin{equation*}
l_{AB}^{a}=l_{r}^{\prime 0}r_{0}^{\prime }+l_{r}^{\prime 1}r_{1}^{\prime
}=-Kl_{0}r_{0}^{\prime }+(1+\beta _{r})(1/K)l_{0}r_{1}^{\prime }.
\end{equation*}%
If only spatial parts of $l_{r}^{\mu }$ and $l_{r}^{\prime \mu }$ are
compared than one finds that $\infty \succ l_{r}^{\prime 1}\geq l_{0}$ for $%
-1/2\prec \beta _{r}\leq 0$ and $l_{0}\leq l_{r}^{\prime 1}\prec \infty $
for $0\leq \beta _{r}\prec \infty $ , which once again shows that such
comparison is physically meaningless in the \textquotedblleft TT
relativity.\textquotedblright\ Hovewer the invariant spacetime length always
takes the same value
\begin{equation*}
l=l_{r}^{\prime }=(l_{r}^{^{\prime }\mu }l_{\mu r}^{\prime
})^{1/2}=l_{r}=l_{0},
\end{equation*}%
and as already said, it can be compared in a physically meaningful sense in
the \textquotedblleft TT relativity.\textquotedblright\ One concludes from
this discussion that, e.g., our particular 4-vector $l_{AB}^{a}$ (a
geometrical quantity) is represented in different bases $\left\{ e_{\mu
}\right\} $, $\left\{ e_{\mu }^{\prime }\right\} $, $\left\{ r_{\mu
}\right\} $ and $\left\{ r_{\mu }^{\prime }\right\} $ by its coordinate
representations $l_{e}^{\mu }$, $l_{e}^{\prime \mu }$, $l_{r}^{\mu }$ and $%
l_{r}^{\prime \mu }$, respectively$.$

We see that in the \textquotedblleft TT relativity\textquotedblright\ the
geometrical quantities, e.g., the 4-vectors $x^{a}$, $l^{a}$,..., have
different representations depending on the chosen IFR and the chosen
coordinatization in that IFR, e.g., $x_{e,r}^{\mu }$, $l_{e,r}^{\prime \mu }$%
, ... . Although the Einstein coordinatization is preferred by physicists
due to its simplicity and symmetry it is nothing more \textquotedblleft
physical\textquotedblright\ than others, e.g., the \textquotedblleft
r\textquotedblright\ coordinatization. The coordinate dependent quantities
have not an intrinsic physical meaning. The spacetime length $l$ is an
example of a well defined quantity that is independent of the chosen IFR and
also of the coordinatization taken in that IFR; it is an intrinsic property
of the spacetime. From this consideration an important conclusion emerges;
the usual 3D length of a moving object cannot be defined in the 4D spacetime
of the TT relativity in an adequate way, since it is only the spatial length
and not a 4D tensor quantity.\newline
\newline
\newline
{2.2. \textbf{The AT of Length}}\newline
\newline
In contrast to the covariant definition of the spacetime length and the TT
of the spacetime tensors considered in the \textquotedblleft TT
relativity\textquotedblright\ \textit{the synchronous definition of spatial
length, introduced by Einstein} [12] \textit{defines length as the spatial
distance between two spatial points on the (moving) object measured by
simultaneity in the rest frame of the observer.} To see the difference with
respect to the \textquotedblleft TT relativity\textquotedblright\ we
determine the spatial length of the rod considered in the previous section.
As shown above in the \textquotedblleft TT relativity,\textquotedblright\ in
contrast to the \textquotedblleft AT relativity,\textquotedblright\ one
cannot speak about the spatial distance, as a correctly defined physical
quantity, but only about 4D tensor quantities, the geometrical quantities -
the position 4-vectors $x_{A,B}^{a}$, the distance 4-vector $l_{AB}^{a}$,
the spacetime length $l$, etc., and their 4D representations, $%
x_{A,B,..e,r,..}^{\mu }$, $l_{ABe,r,..}^{\mu }$, $l_{e,r,..}$.

Instead of to work with geometrical quantities $x_{A,B}^{a}$, $l_{AB}^{a}$
and $l$ one deals, in the \textquotedblleft AT
relativity,\textquotedblright\ only with the spatial, or temporal, parts of
their coordinate representations $x_{Ae,r}^{\mu }$, $x_{Be,r}^{\mu }$ and $%
l_{e,r}^{\mu }$. First the \textquotedblleft e\textquotedblright\
coordinatization, which is almost always used in the \textquotedblleft AT
relativity,\textquotedblright\ is considered. According to Einstein's
definition [12] of the spatial length the spatial ends of the rod must be
taken simultaneously in the chosen coordinatization. In 4D (at us 2D)
spacetime and in the \textquotedblleft e\textquotedblright\ coordinatization
the simultaneous events $A$ and $B$ (whose spatial parts correspond to the
spatial ends of the rod) are the intersections of $x_{e}^{1}$ axis (that is
along the spatial base vector $e_{1}$) and the world lines of the spatial
ends of the rod that is at rest in $S$ and situated along the $x_{e}^{1}$
axis. The position 4-vectors (in the \textquotedblleft e\textquotedblright\
base) $x_{Ae}^{\mu }$ and $x_{Be}^{\mu }$ of the simultaneous (at $%
t_{e}=a=0) $ events $A$ and $B$ in $S$ are $x_{Ae}^{\mu }=0e_{0}+0e_{1}$,
or, in short, $x_{Ae}^{\mu }=(0,0)$, and $x_{Be}^{\mu }=(0,l_{0})$, and the
distance 4-vector (in the \textquotedblleft e\textquotedblright\ base) $%
l_{ABe}^{\mu }=x_{Be}^{\mu }-x_{Ae}^{\mu }=(0,l_{0})$. We emphasize that it
is necessary in the \textquotedblleft AT relativity\textquotedblright\ to
take the end points of the spatial length of the rod to be simultaneous,
whereas in the \textquotedblleft TT relativity\textquotedblright\ the events
$A$ and $B$ can be, in principle, taken at arbitrary $x_{Ae}^{0}\neq
x_{Be}^{0}$. Then in $S$, the rest frame of the object, the spatial part $%
l_{ABe}^{1}=l_{0}$ of $l_{ABe}^{\mu }$ is considered to define the rest
spatial length (the temporal part of $l_{ABe}^{\mu }$ is taken to be zero).
Further one uses the inverse Lorentz transformations to express $x_{Ae}^{\mu
}$, $x_{Be}^{\mu }$, and $l_{ABe}^{\mu }$ in $S$ in terms of the
corresponding quantities in $S^{\prime }$, in which the rod is moving. This
procedure yields $x_{A,Be}^{0}=ct_{A,Be}=\gamma _{e}(ct_{A,Be}^{\prime
}+\beta _{e}x_{A,Be}^{\prime 1})$, and $x_{A,Be}^{1}=\gamma _{e}(\beta
_{e}ct_{A,Be}^{\prime }+x_{A,Be}^{\prime 1}$, whence
\begin{equation}
l_{ABe}^{0}=ct_{Be}-ct_{Ae}=\gamma _{e}(ct_{Be}^{\prime }-ct_{Ae}^{\prime
})+\gamma _{e}\beta _{e}(x_{Be}^{\prime 1}-x_{Ae}^{\prime 1})=\gamma
_{e}l_{ABe}^{\prime 0}+\gamma _{e}\beta _{e}l_{ABe}^{\prime 1}
\label{elnula}
\end{equation}%
and
\begin{equation}
l_{ABe}^{1}=x_{Be}^{1}-x_{Ae}^{1}=\gamma _{e}(x_{Be}^{\prime
1}-x_{Ae}^{\prime 1})+\gamma _{e}\beta _{e}(ct_{Be}^{\prime
}-ct_{Ae}^{\prime })=\gamma _{e}l_{ABe}^{\prime 1}+\gamma _{e}\beta
_{e}l_{ABe}^{\prime 0}.  \label{eljedan}
\end{equation}%
Now comes the main difference between the two forms of relativity. Instead
of to work with 4D tensor quantities and their LT (as in the
\textquotedblleft TT relativity\textquotedblright ) in the \textquotedblleft
AT relativity\textquotedblright\ one forgets about the transformation of the
temporal part $l_{ABe}^{0}$, Eq.(\ref{elnula}), and considers only the
transformation of the spatial part $l_{ABe}^{1}$, Eq.(\ref{eljedan}).
Further, in that relation for $l_{ABe}^{1}$ one assumes that $t_{Be}^{\prime
}=t_{Ae}^{\prime }=t_{e}^{\prime }=b$, i.e., that $x_{Be}^{\prime 1}$ and $%
x_{Ae}^{\prime 1}$ are simultaneously determined at some arbitrary $%
t_{e}^{\prime }=b$ in $S^{\prime }$. However, in 4D (at us 2D) spacetime
such an assumption means that in $S^{\prime }$ one actually \emph{does not
consider the same events} $A$ and $B$ as in $S$ but some other two events $C$
and $D$, whence $t_{Be}^{\prime }=t_{Ae}^{\prime }$ has to be replaced with $%
t_{De}^{\prime }=t_{Ce}^{\prime }=b$. The events $C$ and $D$ are the
intersections of the line (the hypersurface $t_{e}^{\prime }=b$ with
arbitrary $b$) parallel to the spatial axis $x_{e}^{\prime 1}$ (which is
along the spatial base vector $e_{1}^{\prime }$) and of the above mentioned
world lines of the spatial end points of the rod. Then in the above
transformation for $l_{ABe}^{1}$ (\ref{eljedan}) one has to write $%
x_{De}^{\prime 1}-x_{Ce}^{\prime 1}=l_{CDe}^{\prime 1}$ instead of $%
x_{Be}^{\prime 1}-x_{Ae}^{\prime 1}=l_{ABe}^{\prime 1}$. The spatial parts $%
l_{ABe}^{1}$ and $l_{CDe}^{\prime 1}$ are the \textit{spatial distances}
between the events $A$, $B$ and $C$, $D$, respectively. \textit{The spatial
distance }$l_{ABe}^{1}=x_{Be}^{1}-x_{Ae}^{1}$ \textit{defines in the
\textquotedblleft AT relativity,\textquotedblright } \textit{and in the
\textquotedblleft e\textquotedblright\ base}, \textit{the spatial length of
the rod} \textit{at rest in} $S$, while $l_{CDe}^{\prime 1}=x_{De}^{\prime
1}-x_{Ce}^{\prime 1}$ \textit{is considered in the \textquotedblleft AT
relativity,\textquotedblright\ and in the \textquotedblleft
e\textquotedblright\ base, to define} \textit{the spatial length of the
moving rod} \textit{in }$S^{\prime }$. With these definitions we find from
the equation for $l_{ABe}^{1}$ (\ref{eljedan}) the relation between $%
l_{e}^{\prime 1}=l_{CDe}^{\prime 1}$ and $l_{e}^{1}=l_{ABe}^{1}=l_{0}$ as
the famous formulae for the Lorentz contraction of the moving rod
\begin{equation}
l_{e}^{\prime 1}=x_{De}^{\prime 1}-x_{Ce}^{\prime 1}=l_{0}/\gamma
_{e}=(x_{Be}^{1}-x_{Ae}^{1})(1-\beta _{e}^{2})^{1/2},  \label{contr}
\end{equation}%
with $t_{Ce}^{\prime }=t_{De}^{\prime }$, $\,$and $t_{Be}=t_{Ae}$, where $%
\beta _{e}=V_{e}/c$, $V_{e}$ is the relative velocity of $S$ and $S^{\prime
} $. Note that the spatial lengths $l_{0}$ and $l_{e}^{\prime 1}$ refer not
to the same 4D tensor quantity, as in the \textquotedblleft TT
relativity,\textquotedblright\ but to two different quantities in 4D
spacetime. These quantities are obtained by the same measurements in $S$ and
$S^{\prime }$; the spatial ends of the rod are measured simultaneously at
some $t_{e}=a$ in $S$ and also at some $t_{e}^{\prime }=b$ in $S^{\prime }$,
and $a$ in $S$ and $b$ in $S^{\prime }$ are not related by the LT or any
other coordinate transformation. While in the \textquotedblleft TT
relativity\textquotedblright\ one deals with events as correctly defined
quantities in 4D spacetime in Einstein's approach [12] the spatial and
temporal parts of events are treated separately, and moreover the time
component is not transformed in the Lorentz contraction.

The LT (\ref{fah}) is the transformation in 4D spacetime and it transforms
some 4D tensor quantity $Q_{b..}^{a..}(x^{c},x^{d},..)$ from $S$ to $%
Q_{b..}^{\prime a..}(x^{\prime c},x^{\prime d},..)$ in $S^{\prime }$, (all
parts of the quantity are transformed), which means that in 4D spacetime is
not possible to neglect the transformation of $l^{0}$ as a part of $l^{\mu }$%
, as done in the derivation of the Lorentz contraction (\ref{contr}).
However, if one does not forget the transformation of the temporal part $%
l_{ABe}^{0}$, Eq.(\ref{elnula}), and takes in it that $t_{Be}^{\prime
}=t_{Ae}^{\prime }$, $t_{Be}=t_{Ae}$ (as in the derivation of the Lorentz
contraction), then one finds from (\ref{elnula}) that $x_{Be}^{\prime
1}=x_{Ae}^{\prime 1}$, which is in the obvious contrast with the formulae
for the Lorentz contraction.

Let us also see does the Lorentz contraction, as the coordinate
transformation, change the interval $ds$, which defines the geometry of the
spacetime. In $S$ and in the \textquotedblleft e\textquotedblright\ base the
interval $ds$ is $ds^{2}=ds_{e}^{2}=(dx_{e}^{1})^{2}-(c^{2}dt_{e})^{2}$, and
with $dt_{e}=0$, as assumed in the derivation of the Lorentz contraction, it
becomes, in $S\quad ds^{2}=(dx_{e}^{1})^{2}$. In $S^{\prime }$, where it is
assumed that $dt_{e}^{\prime }=0$, and with the relation for the Lorentz
contraction (\ref{contr}), $dx_{e}^{\prime 1}=dx_{e}^{1}/\gamma _{e}$, the
infinitesimal spacetime distance $ds^{\prime }$ becomes, in $S^{\prime }$%
\quad $ds^{\prime 2}=(dx_{e}^{1})^{2}/\gamma _{e}^{2}$, and thus $ds^{\prime
}\neq ds$.

Let us now consider the Lorentz \textquotedblleft
contraction\textquotedblright\ in the \textquotedblleft r\textquotedblright\
coordinatization. According to Einstein's definition [12] of the spatial
length the spatial ends of the rod must be taken simultaneously in the
chosen coordinatization. In 4D (at us 2D) spacetime and in the
\textquotedblleft r\textquotedblright\ base the spatial ends of the
considered rod, that is at rest in $S$, must lie on the light line, i.e., on
the $x_{r}^{1}$ axis (that is along the spatial base vector $r_{1}$). Hence
the simultaneous events $E$ and $F$ (whose spatial parts correspond to the
spatial ends of the rod) are the intersections of $x_{r}^{1}$ axis and the
world lines of the spatial ends of the rod. Note that in our 2D spacetime
the events $E$ and $F$ are not the same events as the events $A$ and $B$,
considered in the \textquotedblleft e\textquotedblright\ base for the same
rod at rest in $S$, since the simultaneity of the events is defined in
different ways. The $\left\{ r_{\mu }\right\} $ representations of the
position 4-vectors $x_{E}^{a}$ and $x_{F}^{a}$ of the events $E$ and $F$ in $%
S$ are $x_{Er}^{\mu }=(0,0)$ and $x_{Fr}^{\mu }=(0,l_{0})$, and of the
distance 4-vector $l_{EF}^{a}$ is $l_{EFr}^{\mu }=x_{Fr}^{\mu }-x_{Er}^{\mu
}=(0,l_{r}^{1})=(0,l_{0})$. However, as noticed above, in 4D spacetime the
spatial length in the \textquotedblleft r\textquotedblright\ base $%
l_{r}^{1}=l_{0}$ (with $x_{Fr}^{0}=x_{Er}^{0}$) is not the same 4D quantity
as the spatial length in the \textquotedblleft e\textquotedblright\ base $%
l_{e}^{1}=l_{0}$ (with $x_{Be}^{0}=x_{Ae}^{0})$, since the simultaneity is
defined in a different way. Applying the same procedure as in the case of
the derivation of the Lorentz contraction in the \textquotedblleft
e\textquotedblright\ base we find the relations for $l_{r}^{0}$ and $%
l_{r}^{1}$ corresponding to (\ref{elnula}) and (\ref{eljedan}),
respectively,
\begin{equation}
l_{EFr}^{0}=x_{Fr}^{0}-x_{Er}^{0}=(1/K)(x_{Fr}^{\prime 0}-x_{Er}^{\prime
0})=(1/K)l_{EFr}^{\prime 0},  \label{elnul1}
\end{equation}%
\begin{equation}
l_{EFr}^{1}=x_{Fr}^{1}-x_{Er}^{1}=(\beta _{r}/K)(x_{Fr}^{\prime
0}-x_{Er}^{\prime 0})+K(x_{Fr}^{\prime 1}-x_{Er}^{\prime 1})=(\beta
_{r}/K)l_{EFr}^{\prime 0}+Kl_{EFr}^{\prime 1},  \label{eljed1}
\end{equation}%
$K=(1+2\beta _{r})^{1/2}$. Further, in the \textquotedblleft
r\textquotedblright\ base, one again forgets the transformation of the
temporal part $l_{EFr}^{0}$ (\ref{elnul1}) of $l_{EFr}^{\mu }$ and assumes
that in the relation for $l_{EFr}^{1}$ (\ref{eljed1}) $x_{Fr}^{\prime 1}$
and $x_{Er}^{\prime 1}$ are simultaneously determined at some $%
x_{Fr}^{\prime 0}=x_{Er}^{\prime 0}=b$ in $S^{\prime }$. However, in the
same way as in the \textquotedblleft e\textquotedblright\ base, in 4D (at us
2D) spacetime such an assumption means that in $S^{\prime }$ one actually
\emph{does not consider the same events} $E$ and $F$ as in $S$ but some
other two events $G$ and $H$, and that the equality $x_{Fr}^{\prime
0}=x_{Er}^{\prime 0}=b$ has to be replaced by $x_{Hr}^{\prime
0}=x_{Gr}^{\prime 0}=b$. The events $G$ and $H$ are the intersections of the
line (the hypersurface $x_{Hr}^{\prime 0}=x_{Gr}^{\prime 0}=b$ with
arbitrary $b$) parallel to the spatial axis $x_{r}^{\prime 1}$ (which is
along the spatial base vector $r_{1}^{\prime }$) and of the above mentioned
world lines of the spatial end points of the rod. Then in the above
transformation for $l_{EFr}^{1}$ (\ref{eljed1}) $l_{GHr}^{\prime 0}=0$, and $%
l_{EFr}^{\prime 1}=x_{Fr}^{\prime 1}-x_{Er}^{\prime 1}$ is replaced by $%
l_{GHr}^{\prime 1}=x_{Hr}^{\prime 1}-x_{Gr}^{\prime 1}$. Now, \textit{in the
\textquotedblleft r\textquotedblright\ base}, \textit{the spatial distance }$%
l_{EFr}^{1}=x_{Fr}^{1}-x_{Er}^{1}$ \textit{defines in the \textquotedblleft
AT relativity\textquotedblright } \textit{the spatial length of the rod}
\textit{at rest in} $S$, while $l_{GHr}^{\prime 1}=x_{Hr}^{\prime
1}-x_{Gr}^{\prime 1}$ \textit{defines} \textit{the spatial length of the
moving rod} \textit{in }$S^{\prime }$. Then, from the equation for $%
l_{EFr}^{1}$ (\ref{eljed1}), and with these definitions, we find the
relation between $l_{r}^{\prime 1}=l_{GHr}^{\prime 1}$ and $%
l_{r}^{1}=l_{EFr}^{1}=l_{0}$ as the Lorentz \textquotedblleft
contraction\textquotedblright\ of the moving rod in the \textquotedblleft
r\textquotedblright\ base,
\begin{equation}
l_{r}^{\prime 1}=x_{Hr}^{\prime 1}-x_{Gr}^{\prime
1}=l_{0}/K=(1/K)(x_{Fr}^{1}-x_{Er}^{1}),  \label{contra1}
\end{equation}%
with $\,x_{Hr}^{\prime 0}=x_{Gr}^{\prime 0}$ $\,$and $x_{Fr}^{0}=x_{Er}^{0}$%
. In contrast to the \textquotedblleft e\textquotedblright\ coordinatization
we find that in the \textquotedblleft r\textquotedblright\ base there is a
length dilatation $\infty \succ l_{r}^{\prime 1}\succ l_{0}$ for $-1/2\prec
\beta _{r}\prec 0$ and the standard \textquotedblleft length
contraction\textquotedblright\ $l_{0}\succ l_{r}^{\prime 1}\succ 0$ for
positive $\beta _{r}$, which clearly shows that the \textquotedblleft
Lorentz contraction\textquotedblright\ is not physically correctly defined
transformation.

We see from the preceding discussion that - \textit{the Lorentz contraction
is the transformation connecting different quantities (in 4D spacetime) in
different IFRs and different coordinatizations, and also it changes the
infinitesimal spacetime distance }$ds$\textit{\ and consequently the
pseudo-Euclidean geometry of the 4D spacetime. Such characteristics of the
Lorentz contraction as the coordinate transformation clearly show that the
Lorentz contraction belongs to - the AT. }In the same way\textit{\ }one can
show that the usual \textquotedblleft time dilatation\textquotedblright\
does have the same characteristics as the Lorentz contraction, i.e., that it
is also - an AT, but this will not be done here.

Although the Lorentz contraction is an AT it is still widely used in
numerous textbooks and papers as an \textquotedblleft important relativistic
effect.\textquotedblright\ Thus, for example, it is almost generally
accepted in ultra-relativistic nuclear collisions, see, e.g., [13]:
\textquotedblleft that in the center-of-mass frame two highly Lorentz
contracted nuclei pass through each other .... .\textquotedblright\ In the
recent paper [14] it is supposed that: \textquotedblleft ... \v{C}erenkov
radiation of the charged two-particle system involves the Lorentz
contraction of their rest distance.\textquotedblright\ An experiment, based
on this idea, is suggested in [14] for the verification of the Lorentz
contraction. Moreover it is argued in [15] that the authors have
experimentally succeeded to observe the Lorentz contraction of magnetic flux
quanta (vortices) in Josephson tunnel junction. In all these examples it is
understood that a Lorentz boost transforms the rest length to the contracted
length. But, as it is already explained above, a Lorentz boost is a TT
transforming from an IFR $S$ to another IFR $S^{\prime }$, e.g., all four
coordinates as a 4-vector; the same events are considered in $S$ and $%
S^{\prime }$. Also, a Lorentz boost transforms a physical quantity
represented by a 4D spacetime tensor, e.g., $Q(x)$ in $S$, to $Q^{\prime
}(x^{\prime })$ in $S^{\prime }$, thus again considering the same quantity
in $S$ and $S^{\prime }$. On the contrary, as already said, in the Lorentz
contraction, Eqs.(\ref{contr}) and (\ref{contra1}), the time component is
not transformed and the Lorentz contraction is an AT from the relativity
viewpoint, which has nothing in common with a Lorentz boost as a TT. In the
\textquotedblleft TT relativity\textquotedblright\ one cannot say that the
nucleus must contract (as argued in literature on ultra-relativistic nuclear
collisions), or that the rest distance between two charged particles
undergoes the Lorentz contraction (as considered in [14]), or that the
vortices contract when moving (as argued to be proved in experiments [15]),
since the Lorentz contraction is certainly not a relativistic relation,
i.e., the relation belonging to the \textquotedblleft TT
relativity,\textquotedblright\ and cannot be used either to illustrate or to
test any part of the \textquotedblleft TT relativity.\textquotedblright

The above discussion reveals the main differences between \textit{the
spacetime length} considered in the \textquotedblleft TT
relativity\textquotedblright\ and \textit{the spatial length} considered in
the \textquotedblleft AT relativity.\textquotedblright\ The same example (a
rod at rest in $S$) is investigated in the \textquotedblleft TT
relativity,\textquotedblright\ Sec.2.1, and in the \textquotedblleft AT
relativity,\textquotedblright\ this section. The \textquotedblleft TT
relativity\textquotedblright\ deals with 4D quantities in 4D spacetime. We
associate with the mentioned rod a 4D quantity - a distance 4-vector $%
l_{AB}^{a}$, and consider this quantity $l_{AB}^{a}$ in two IFRs $S$ and $%
S^{\prime }$ (in which the rod is moving) and in two coordinatizations,
\textquotedblleft e\textquotedblright\ and \textquotedblleft
r\textquotedblright . Four different decompositions, representations, are
found for the same $l_{AB}^{a}$. Different representations of $l_{AB}^{a}$
in $S$ and $S^{\prime }$ are connected by the TT - the LT. In terms of $%
l_{AB}^{a}$ \textit{the spacetime length} $l$ (\ref{covlen}) is constructed
and it \textit{does have the same value for all four representations of} $%
l_{AB}^{a}.$ An essentially different treatment of that rod is performed in
the \textquotedblleft AT relativity.\textquotedblright\ This form of
relativity does not deal with 4D quantities in 4D spacetime. In the
\textquotedblleft AT relativity\textquotedblright\ we associate with that
rod \textit{four different spatial lengths, i.e., four different 3D
quantities in 4D spacetime}; in the \textquotedblleft e\textquotedblright\
base they are $l_{ABe}^{1}$ in $S$ and $l_{CDe}^{\prime 1}$ in $S^{\prime }$%
, and in the \textquotedblleft r\textquotedblright\ base they are $%
l_{EFr}^{1}$ in $S$ and $l_{GHr}^{\prime 1}$ in $S^{\prime }$. The
quantities in the same base but in different IFRs are connected by the AT -
the Lorentz \textquotedblleft contraction\textquotedblright ; $l_{ABe}^{1}$
and $l_{CDe}^{\prime 1}$ with (\ref{contr}), and $l_{EFr}^{1}$ and $%
l_{GHr}^{\prime 1}$ with (\ref{contra1}). None of these quantities is well
defined in 4D spacetime. We conclude from the whole previous consideration
that \textit{when the 4D structure of our spacetime is correctly taken into
account then there is no place for the Lorentz contraction formulae, and
only the spacetime length and the spacetime quantities are well defined
quantities. }\newline
\newline
\newline
\textbf{2.3. \textquotedblleft Car and Garage Paradox\textquotedblright }%
\newline
\newline
In the previous sections we have examined the main characteristics of both
forms of relativity. Now we want to show the difference between the
treatments of relatively moving systems in the \textquotedblleft AT
relativity\textquotedblright\ and the \textquotedblleft TT
relativity.\textquotedblright\ Usually such systems are treated in a
noncovariant manner, i.e., in the \textquotedblleft AT
relativity,\textquotedblright\ but here we shall present the treatment of
such systems in a manifestly covariant manner, i.e., in the
\textquotedblleft TT relativity.\textquotedblright\ In order to see the
differences between both treatments we do not need to work completely in
geometrical terms, but we can choose some specific coordinatization, e.g.,
the simplest one, the \textquotedblright e\textquotedblright\
coordinatization. Therefore, in the following, we restrict ourselves to the
\textquotedblleft e\textquotedblright\ coordinatization and, for simplicity
in notation, we omit the subscript -e- in all quantities. However, we shall
often write the important relations in geometrical terms, and also we shall
explain which results and conclusions are independent of the chosen
coordinatization.

As already discussed, if in an IFR $S^{\prime }$ in which the time component
of the distance 4-vector $l_{AB}^{\prime \mu }$, i.e., $l_{AB}^{\prime 0},$
is zero (simultaneously determined events $A$ and $B$), then $l_{AB}^{\prime
\mu }$ comprises only spatial components. Hence, in $S^{\prime }$ the
invariant spacetime length $l^{\prime }$ is given as the usual 3D distance
between $A$ and $B$. But, in such a case, $l_{AB}^{\mu }$ in the rest frame $%
S$ of the object does have $l_{AB}^{0}\neq 0$, and the spacetime length $l$
in $S$ (it is $=l^{\prime }$) could take different values depending on the
chosen IFR $S^{\prime }$. Such an arbitrariness in $l$, although not
forbidden by any physical law, would complicate both theory and experiment.
Furhermore, one wants to retain the connection with the prerelativistic
concept of the spatial length. Therefore, the most convenient choice for the
frame in which the time component $l_{AB}^{\prime 0}$ of $l_{AB}^{\prime \mu
}$ is zero is the $S$ frame. Thus in $S$ the position 4-vectors $%
x_{A,B}^{\mu }$ are determined simultaneously, $%
x_{B}^{0}-x_{A}^{0}=l_{AB}^{0}=0,$ and the spacetime length $l$ becomes in $%
S $, the rest frame of the object, the rest spatial length $l_{0}$, i.e., $%
l=(l_{AB}^{i}l_{ABi})^{1/2}=l_{0}$, as in the prerelativistic physics. The
observers in all other IFRs will look at the same events but associating
with them different coordinates; they all find (measure) the same value $%
l=l_{0}$ for the spacetime length. Note, as we have mentioned, that if $%
l_{AB}^{0}=0$ then $l_{AB}^{\prime \mu }$ in any other IFR $S^{\prime }$
will contain the time component $l_{AB}^{\prime 0}\neq 0$. We once again
emphasize that the choice of the rest frame of the object as the starting
frame for the consideration is not dictated by physical requirements. This
choice is, in fact, determined only by our desire to have a quantity that
corresponds in the 4D spacetime to the prerelativistic spatial length. As
already said at the end of Sec.2.1 the usual 3D length of a moving object
cannot be defined in the 4D spacetime, i.e., in the \textquotedblleft TT
relativity\textquotedblright , in an adequate way. \emph{Only the spacetime
length }$l$\emph{\ does have a definite theoretical and experimental meaning
and it is an invariant quantity.} This holds for all possible
synchronizations. With our choice of the rest frame of the object as the
starting frame, i.e., with $l=l_{0},$ the spatial rest length determined
simultaneously in $S$ obtains the properties of the spacetime length $l.$
Then, the coordinate measurements of $x_{A,B}^{\prime \mu }$ in an IFR $%
S^{\prime }$ in which an object is moving are not of interest in their own
right but they have to enable one to find the rest spatial length $l_{0}$.
In the prerelativistic \textquotedblleft 3+1\textquotedblright\ picture, and
in the \textquotedblleft AT relativity,\textquotedblright\ one can compare
\textit{the spatial lengths} of two relatively moving objects. But in the
\textquotedblleft TT relativity\textquotedblright\ \textit{the spacetime
lengths} (that contain both spatial and temporal parts) are well defined
quantities in 4D spacetime and they, or the rest spatial lengths, can be
compared in a physically meaningful way.

Let us illustrate the preceding discussion considering the well-known
\textquotedblleft Car and garage paradox\textquotedblright\ (see, e.g., [7],
p.9). The common assertion about this \textquotedblleft
paradox\textquotedblright\ is that it comes out due to, [7]:
\textquotedblright The lack of a notion of absolute simultaneity in special
relativity ....,\textquotedblright\ and consequently due to the relativity
of the Lorentz contraction. However, as discussed above, the relativity of
simultaneity is a coordinate dependent effect and, for example, for $%
\varepsilon =0$, [8], the absolute simultaneity is preserved. Also, the
Lorentz contraction is an AT and it has nothing to do with the 4D
pseudo-Euclidean geometry of the special relativity. Therefore, we discuss
this \textquotedblleft paradox\textquotedblright\ using covariant 4D
quantities, i.e., in the \textquotedblleft TT relativity\textquotedblright .
But for our purposes, as it is already explained, there is no need to
discuss the \textquotedblleft paradox\textquotedblright\ in geometrical
terms, than it can be considered in the inertial coordinate systems with the
Minkowski metric tensors, that is in the \textquotedblleft
e\textquotedblright\ coordinatization. The frame in which a garage is at
rest is denoted by $S$ while that one in which a car is at rest by $%
S^{\prime }$. The unprimed quantities are in $S$ and the primed ones in $%
S^{\prime }$. Instead of 4D spacetime we work here with 2D spacetime. The
garage and the car are along the common $x^{1},x^{\prime 1}$ - axes, and
they are of equal proper lengths $l_{g}=l_{c}^{\prime }=l$. First we
consider the \textquotedblleft paradox\textquotedblright\ in the $S$ frame.
The distance 4-vector $l_{g}^{\mu }$ for the garage is determined directly
in $S$. However the distance 4-vector $l_{c}^{\mu }$ for the moving car in $%
S $ has to be determined in such a way that one first finds it in the car
own rest frame $S^{\prime }$, and then transforms it by the LT to the rest
frame of the garage $S$. This procedure follows from the preceding general
discussion where it is shown that only the rest length, i.e., the invariant
spacetime length, of a moving object is well defined quantity in the
\textquotedblleft TT relativity\textquotedblright . The position 4-vectors
of the spatial end points $A$ and $B$ of the garage in $S$ are taken to be $%
x_{Ag}^{\mu }=(0,0)$ and $x_{Bg}^{\mu }=(0,l)$, and the distance 4-vector is
$l_{g}^{\mu }=x_{Bg}^{\mu }-x_{Ag}^{\mu }=(0,l)$. (In the remaining part of
this paper we shall use $A$, $B$, $C$, $D$, ... and $A^{\prime }$, $%
B^{\prime }$, ... to denote the spatial points on the (moving) object.) The
invariant spacetime length of the garage is $l_{g}=(l_{g}^{\mu }l_{\mu
g})^{1/2}=l$. The frame $S^{\prime }$, together with the car, moves relative
to $S$ with 4-velocity $v^{\mu }=(\gamma c,\gamma V)$, $\gamma =(1-\beta
^{2})^{1/2}$, $\beta =V/c$. Let the origin of $S^{\prime }$ (with the left
end of the car $A^{\prime }$ attached to it) coincides with the origin of $S$
(with the left end of the garage $A$ attached to it) at the moment $%
t=t^{\prime }=0$. The position 4-vectors of the end points $A^{\prime }$ and
$B^{\prime }$ of the car in $S^{\prime }$ are $x_{A^{\prime }c}^{\prime \mu
}=(0,0)$ and $x_{B^{\prime }c}^{\prime \mu }=(0,l)$, and the distance
4-vector is $l_{c}^{\prime \mu }=x_{B^{\prime }c}^{\prime \mu }-x_{A^{\prime
}c}^{\prime \mu }=(0,l)$. The invariant spacetime length of the car is $%
l_{c}^{\prime }=(l_{c}^{^{\prime }\mu }l_{\mu c}^{\prime })^{1/2}=l$. To
find the position 4-vector of the right end $B^{\prime }$ of the car and the
distance 4-vector of the car in $S$ one applies the LT to $x_{B^{\prime
}c}^{\prime \mu }$ and $l_{c}^{\prime \mu }$. Then one finds $x_{B^{\prime
}c}^{\mu }=l_{c}^{\mu }=(\gamma \beta l,\gamma l)$. The spacetime length of
the car is now determined in $S$, and is as before $l_{c}=l$, equal to its
rest length. Obviously, in the \textquotedblleft TT
relativity\textquotedblright\ nothing could be said whether the \textit{%
moving }car fit into the garage or not, as the comparison of the spatial
parts of the distance 4-vectors for the garage and the car has no physical
meaning. (Note that there is a \textit{dilatation }of the spatial part of
the moving car, and not the Lorentz contraction as in the \textquotedblleft
AT relativity\textquotedblright .) Only if both objects are relatively at
rest their spatial parts can be compared in physically meaningful sense. The
same conclusions hold from the point of view of the observer in $S^{\prime }$%
, or any other IFR. These conclusions hold not only for Einstein's
coordinatization (which is used here) but for any other possible
coordinatization of IFRs.\newline
\newline
\newline
\textbf{3. CURRENT-CARRYING CONDUCTOR AND EXTERNAL}

\textbf{ELECTRIC FIELDS IN THE \textquotedblleft TT
RELATIVITY\textquotedblright }\newline
\newline
Let us now apply these ideas to the consideration of a CCC in the
\textquotedblleft TT relativity\textquotedblright . An infinite straight
wire with a steady current is situated along the $x^{1}$ axis. A current is
flowing in $-x^{1}$ direction and accordingly electrons move in $+x^{1}$
direction. We suppose that positive and negative charge densities are of
equal magnitude when both subsystems are relatively at rest, i.e., before a
current is established in the wire. In a CCC the wire (i.e., the ions) is
supposed to be at rest in $S$, while the electrons are at rest in $S^{\prime
}$.

Before determining the current density 4-vectors $j^{\mu }$ in $S$ and $%
S^{\prime }$ we give the manifestly covariant definition of a charge within
a boundary $\delta H$ of an arbitrary hypersurface $H$ (see, e.g., [16,
17]),
\begin{equation}
Q_{\delta H}=\int_{H}j^{\mu }d\sigma _{\mu ,}  \label{charge}
\end{equation}%
where $d\sigma ^{\mu }$ is the 4-vector of an element of the hypersurface $H$%
. (This expression can be written in a more general form, i.e., in
geometrical terms, replacing Greek index $\mu $ by the abstract index $a.$)
The invariance of charge defined by (\ref{charge}) is proved in [16] for a
linear CCC, and also for the general case of 4D spacetime in [17]. If the
hypersurface $H$ is chosen in the rest frame of charges in such a way that
it is the space-like plane $t=const.$, then the charge $dQ$ is given as in
the prerelativistic physics $dQ=\rho dV$; the 4-vector $j^{\mu }$ has only
the time component $j^{0}=c\rho $, since we are in the rest frame of
charges, and $d\sigma _{\mu }$ also has only time component which is the
synchronously defined rest volume $dV$. Consequently, the charge density $%
\rho $ is defined as the ratio of $dQ$ given by (\ref{charge}), but taken
simultaneously in the rest frame of the charges, and the synchronously
defined rest volume $dV$, $\rho =dQ/dV$, and it is well defined quantity
from the \textquotedblleft TT relativity\textquotedblright\ viewpoint.
Obviously the charge density has the common prerelativistic meaning only in
the rest frame of the charges. The charge density of moving charges is not a
well defined quantity from the \textquotedblleft TT
relativity\textquotedblright\ point of view in the same way as the spatial
length or the volume of a moving object are not correctly defined quantities
in the \textquotedblleft TT relativity\textquotedblright . This is in
contrast with the \textquotedblleft AT relativity\textquotedblright\ and the
Lorentz contraction, where the charge density of the moving charges is
defined; it is enhanced by $\gamma =(1-\beta ^{2})^{1/2}$ relative to the
proper charge density due to the Lorentz contraction of the moving volume.
Thereby, when determining the current density 4-vector $j^{\mu }$ in some
IFR in which the charges are moving one first has to find that vector in the
rest frame of the charges, where the space component $\mathbf{j}=0$ and $%
\gamma =1$, and then to transform by the LT so determined $j^{\mu }$ to the
considered IFR. According to this consideration the simplest and the correct
way, from the \textquotedblleft TT relativity\textquotedblright\ viewpoint,
to determine the current density 4-vector $j^{\mu }$ in some IFR for a CCC
is the following: The current density 4-vectors $j_{+}^{\mu }$ and $%
j_{-}^{\mu }$ for positive and negative charges, respectively, have to be
determined in their rest frames and then transformed by the LT to the given
IFR. It has to be noted that in the \textquotedblleft TT
relativity\textquotedblright\ it is not necessary to determine $j^{\mu }$
for a CCC, in an arbitrary IFR, in the mentioned way. In that frame we could
start in (\ref{charge}) with an arbitrary space-like hypersurface $H$ and
determine $j^{a}$ and $d\sigma ^{a}$ in some coordinatization that is
different than the \textquotedblleft e\textquotedblright\ coordinatization.
But then we loose the connection with the prerelativistic notions, the
charge density $\rho $, the current density $\mathbf{j}$ (3-vector), the
spatial length and volume, etc., and with the prerelativistic relation $%
dQ=\rho dV$.\newline
\newline
\newline
\textbf{3.1. The Current Density $j^{\mu }$ in the Ions' Rest Frame S}%
\newline
\newline
Hence, the current density 4-vector in $S$, for the considered wire with
current, is $j^{\mu }=j_{+}^{\mu }+j_{-}^{\mu }$, where $j_{+}^{\mu }=(c\rho
_{0},0)$. The positive charge density $\rho _{+}$ is $=\rho _{0}$, where $%
\rho _{0}$ is the positive charge density for the wire at rest but without a
current. To find $j_{-}^{\mu }$ in $S$ one has, as already said, to find the
electrons' charge density $\rho _{-}^{\prime }$, and the current density
4-vector of the electrons $j_{-}^{\prime \mu }$ in their rest frame $%
S^{\prime }$, where $\rho _{-}^{\prime }$ is well defined quantity, and then
to transform them to the ions' rest frame $S$. In the rest wire, but without
a current, the charge density of the electrons, which are at rest there, is $%
-\rho _{0}$. Then, it follows from the previous consideration that in $%
S^{\prime }$, where the electrons in that wire, but with a current, are at
rest, the proper charge density $\rho _{-}^{\prime }$ of the electrons must
again be equal to $-\rho _{0}$, i.e.,
\begin{equation}
\rho _{-}^{\prime }=-\rho _{0},\quad j_{-}^{\prime \mu }=(-c\rho _{0},0).
\label{ronula}
\end{equation}%
By means of (\ref{ronula}) and the LT we find the current densities in $S$
as
\begin{equation}
j_{-}^{\mu }=(-c\gamma \rho _{0},-c\gamma \beta \rho _{0}),\quad j^{\mu
}=(c(1-\gamma )\rho _{0},-c\gamma \beta \rho _{0}).  \label{jotmi}
\end{equation}%
Eqs. (\ref{ronula}) and (\ref{jotmi}) are in contrast to all previous works
from the time of Clausius, Clausius hypothesis, see [18], until today. In
the Clausius hypothesis \textit{it is simply supposed} that in the ions'
rest frame $S$ the charge density of the moving electrons $\rho _{-}=-\rho
_{0}$. However the same equations were already obtained in [5], where the
\textquotedblleft AT relativity\textquotedblright\ with the Lorentz
contraction is used. This may seem surprising that the same equations exist
in [5] (with the \textquotedblleft AT relativity\textquotedblright ) and
here, where the \textquotedblleft TT relativity\textquotedblright\ is
considered and thus only the covariant quantities are used. But, we must
note that the results obtained in [5] are not actually based on the AT,
i.e., on the Lorentz contraction, than on the \textit{assumption} that in
the electrons' rest frame $S^{\prime }$ the electrons' charge density $\rho
_{-}^{\prime }$ is $=-\rho _{0}$. In a covariant approach, i.e., in the
\textquotedblleft TT relativity,\textquotedblright\ Eq. (\ref{ronula}) is
neither \textit{hypothesis} (as in the traditional approach) nor the \textit{%
assumption} (as in [5]), but it is a consequence of the covariant definition
of an invariant charge (\ref{charge}) and of the invariance of the rest
length, i.e., it resulted from the use of correctly defined covariant
quantities.\newline
\newline
\newline
\textbf{3.2. The $F^{\alpha \beta }$ and the $E^{\alpha },B^{\alpha }$
Formulations of Electrodynamics}\newline
\newline
Having determined the sources $j^{\mu }$ we find the electric and magnetic
fields for that infinite wire with current. One way is to start with the
covariant Maxwell equations with $F^{\alpha \beta }$ and its dual $^{\ast
}F^{\alpha \beta }$
\begin{equation}
\partial _{\alpha }F^{a\beta }=-j^{\beta }/\varepsilon _{0}c,\quad \partial
_{\alpha }\ ^{\ast }F^{\alpha \beta }=0  \label{covef}
\end{equation}%
where $^{\ast }F^{\alpha \beta }=-(1/2)\varepsilon ^{\alpha \beta \gamma
\delta }F_{\gamma \delta }$ and $\varepsilon ^{\alpha \beta \gamma \delta }$
is the totally skew-symmetric Levi-Civita pseudotensor. In such a covariant
formulation F$^{\alpha \beta }$ is the primary quantity; it is the solution
of (\ref{covef}), or the corresponding wave equation
\begin{equation}
\partial ^{\sigma }\partial _{\sigma }F_{\alpha \beta }-(1/\varepsilon
_{0}c)(\partial _{\beta }j_{\alpha }-\partial _{\alpha }j_{\beta })=0,
\label{wave}
\end{equation}%
and it conveys all the information about the electromagnetic field. There is
no need to introduce either the intermediate electromagnetic 4-potential $%
A^{\mu }$ or the connection of the components of $F^{\alpha \beta }$ with
the usual 3-vectors $\mathbf{E}$ and $\mathbf{B.}$ The general solution in
the retarded representation of (\ref{wave}) or (\ref{covef}) is
\begin{equation}
F^{\alpha \beta }(x^{\mu })=(2k/i\pi c)\int \left\{ \frac{\left[ j^{\alpha
}(x^{\prime \mu })(x-x^{\prime })^{\beta }-j^{\beta }(x^{\prime \mu
})(x-x^{\prime })^{\alpha }\right] }{\left[ (x-x^{\prime })^{\sigma
}(x-x^{\prime })_{\sigma }\right] ^{2}}\right\} d^{4}x^{\prime },
\label{elten}
\end{equation}%
where $x^{\alpha }$, $x^{\prime \alpha }$ are the position 4-vectors of the
field point and the source point respectively, $k=1/4\pi \varepsilon _{0}$.
After transforming by the LT (\ref{covef}) to the $S^{\prime }$ frame one
finds the same equations with primed quantities replacing the unprimed ones,
since the transformations of all quantities in (\ref{covef}) are -- the TT.

Instead of to work with $F^{\alpha \beta }$- formulation one can
equivalently use the $E^{\alpha },B^{\alpha }$- formulation, which is
presented in [1] and [3]. It is shown there that in the \textquotedblleft TT
relativity\textquotedblright\ one has to use the 4-vectors $E^{\alpha }$ and
$B^{\alpha }$ instead of the usual 3-vectors $\mathbf{E}$ and $\mathbf{B}.$
The usual transformations of $\mathbf{E}$ and $\mathbf{B}$, obtained by the
identification of the components of $\mathbf{E}$ and $\mathbf{B}$ with the
components of $F^{\alpha \beta }$ ($E_{i}=F^{0i}$ and $B_{i}=^{\ast }F^{0i}$%
), are shown to be the AT referring to the same measurements in different
IFRs and not to the same quantity, see [1] and [3]. In that way it is found
in [1] (and [3]) that contrary to the common belief the usual noncovariant
formulation with the 3-vectors $\mathbf{E}$ and $\mathbf{B}$ is not
equivalent to the covariant formulations. $E^{\alpha }$ and $B^{\alpha }$
are determined by the covariant Maxwell equations derived in [1] (and [3]),
\begin{eqnarray}
\partial _{\alpha }(\delta _{\mu \nu }^{\alpha \beta }v^{\mu }E^{\nu
})+c\partial _{\alpha }(\varepsilon ^{\alpha \beta \mu \nu }B_{\mu }v_{\nu
}) &=&-j^{\beta }/\varepsilon _{0},  \notag \\
\partial _{\alpha }(\delta _{\mu \nu }^{\alpha \beta }v^{\mu }B^{\nu
})+(1/c)\partial _{\alpha }(\varepsilon ^{\alpha \beta \mu \nu }v_{\mu
}E_{\nu }) &=&0,  \label{maeb}
\end{eqnarray}%
where $E^{\alpha }$ and $B^{\alpha }$ are the electric and magnetic field
4-vectors measured by a family of observers moving with 4-velocity $v^{\mu }$%
, and $\delta _{\mu \nu }^{\alpha \beta }=\delta _{\mu }^{\alpha }\delta
_{\nu }^{\beta }-\delta _{\nu }^{\alpha }\delta _{\mu }^{\beta }$. For the
given sources $j^{\mu }$ one could solve these equations and find the
general solutions for $E^{\alpha }$ and $B^{\alpha }$.

We note that it is possible to write Eqs. (\ref{maeb}) in a somewhat simpler
form, the covariant Majorana form, introducing $\Psi ^{\alpha }=E^{\alpha
}-icB^{\alpha }$. Then the covariant Majorana form of Maxwell's equations
becomes
\begin{equation}
(\gamma ^{\mu })^{\beta }\ _{\alpha }\partial _{\mu }\Psi ^{\alpha
}=-j^{\beta }/\varepsilon _{0},  \label{Major}
\end{equation}%
where the $\gamma $-matrices are
\begin{equation}
(\gamma ^{\mu })^{\beta }\ _{\alpha }=\delta _{\rho \gamma }^{\mu \beta
}v^{\rho }g_{\alpha }^{\gamma }+i\varepsilon ^{\mu \beta }\ _{\alpha \gamma
}v^{\gamma }.  \label{gama}
\end{equation}%
In the case that $j^{\mu }=0$ Eq. (\ref{Major}) becomes Dirac-like
relativistic wave equation for free photons
\begin{equation}
(\gamma ^{\mu })^{\beta }\ _{\alpha }\partial _{\mu }\Psi ^{\alpha }=0.
\label{gam1}
\end{equation}%
We shall not further discuss the covariant Majorana formulation since it
will be reported elsewhere.\newline
\newline
\newline
\textbf{3.3. $E^{\alpha }$ for a CCC in the \textquotedblright TT
Relativity\textquotedblright }\newline
\newline
Instead of to solve (\ref{maeb}) or (\ref{Major}) to find $E^{\alpha }$ and $%
B^{\alpha }$ for a CCC we first find $F^{\alpha \beta }$ from (\ref{elten})
inserting into it $j^{\mu }$ from (\ref{jotmi}) and performing the
integration. Then we use the relations given in [1] (and [3]), which connect
$F^{\alpha \beta }$ and $E^{\alpha }$, $B^{\alpha }$- covariant
formulations,
\begin{equation}
E^{\alpha }=(1/c)F^{\alpha \beta }v_{\beta ,}\quad B^{\alpha
}=(1/c^{2})^{\ast }F^{\alpha \beta }v_{\beta }.  \label{vezaEF}
\end{equation}%
The inverse relations connecting the $E^{\alpha },B^{\alpha }$ and $%
F^{\alpha \beta }$- covariant formulations are also given in [1] (and [3])
and they are
\begin{equation}
F^{\alpha \beta }=(1/c)\delta _{\mu \nu }^{\alpha \beta }v^{\mu }E^{\nu
}+\varepsilon ^{\alpha \beta \mu \nu }B_{\mu }v_{\nu },\ ^{\ast }F^{\alpha
\beta }=\delta _{\mu \nu }^{\alpha \beta }v^{\mu }B^{\nu }+(1/c)\varepsilon
^{\alpha \beta \mu \nu }v_{\mu }E_{\nu }.  \label{vezFE}
\end{equation}%
Taking that the family of observers who measures $E^{\alpha }$ is at rest in
the S frame, i.e., that $v_{\mu }=(-c,0)$ one finds from (\ref{vezaEF}) that
$E^{0}=0$,$\ E^{i}=F^{0i}$, whence
\begin{equation}
E^{1}=0,\ E^{2}=2k(1-\gamma )\rho _{0}y(y^{2}+z^{2})^{-1},\
E^{3}=2k(1-\gamma )\rho _{0}z(y^{2}+z^{2})^{-1}.  \label{eovi}
\end{equation}%
\textit{The equation} (\ref{eovi}) \textit{shows that the observer who is at
rest relative to a wire with steady current will see, i.e., measure, the
second order electric field outside such a CCC}. Thus the result which is
for such fields predicted on different grounds in [5] is proved to be
correct in the \textquotedblleft TT relativity\textquotedblright\ too, i.e.,
when all quantities are treated in a covariant manner, see the discussion at
the end of Sec.3.1. We thus find that these electric fields naturally come
out in the \textquotedblleft TT relativity\textquotedblright\ treatment of
physical systems consisting of relatively moving subsystems. Such fields
were not searched for and, it seems, were not observed earlier due to their
extreme smallness. (To be more precise, the similar second-order electric
fields ($\propto v^{2}/c^{2}$) have been detected in [18] and [19], but it
is not sure that they are caused by the effect predicted in [5] and here.)
However, I suppose that such fields must play an important role in many
physical phenomena with steady currents, particularly in tokamaks and
astrophysics, where high currents exist, and in superconductors, where the
electric fields of zeroth order outside CCCs are absent, (see [20]).

Similarly, the magnetic field 4-vector $B^{\mu }$ can be also obtained from (%
\ref{vezaEF}) and the expression for $F^{\alpha \beta }$ (\ref{elten}). For
the observers with $v_{\mu }=(-c,0)$ one finds $B^{0}=0$ and $%
B^{i}=(-1/c)^{\ast }F^{i0}$. In terms of the known $^{\ast }F^{i0}$ we find
for $B^{i}$ the usual expression for the magnetic field of an infinite
straight wire with current, (only the current density is $\gamma $ times
bigger).\newline
\newline
\newline
\textbf{4. CHARGES ON A CURRENT LOOP IN}

\textbf{THE \textquotedblleft TT RELATIVITY\textquotedblright }\newline
\newline
In this section we discuss the macroscopic charge of a square loop with a
steady current in the \textquotedblleft TT relativity\textquotedblright .
This is already discussed in numerous previous works but from the
\textquotedblleft AT relativity\textquotedblright\ viewpoint by using the
synchronous definition of length and the Lorentz contraction. Let a square
loop is at rest in the $x,y$ plane in an IFR $S$. The spatial $x,y$ -
coordinates of the corners are: $A(0,0)$, $B(1,0)$, $E(1,1)$, $F(0,1)$. The
electrons move from $A$ to $B$ in the $\overline{AB}$ side. We first
consider the charge in the $\overline{AB}$ side in two frames; in $S$, the
ions' rest frame, and $S_{AB}^{\prime }$, the electrons' rest frame. The
evaluation of the charge in the $\overline{AB}$ side in $S$ and $%
S_{AB}^{\prime }$ using (\ref{charge}) is already correctly performed in
[16]. However there is an important difference between the calculation of $%
Q_{AB}$ in [16] and in this paper. In [16] the charge density of the
electrons in $S_{AB}^{\prime }$ is not specified but simply taken to have
some undetermined value $-\overline{\lambda }$ and its value in $S$ is found
by means of the LT. The covariance of the definition (\ref{charge}) (the
Lorentz scalar) will yield that $Q_{AB}$ in $S$ is equal to $Q_{AB}^{\prime
} $ in $S_{AB}^{\prime }$ for any choice of $-\overline{\lambda }$. Our
discussion of an infinite wire with a current reveals that $\rho
_{-}^{\prime }$ and $j_{-}^{\prime \mu }$ in $S_{AB}^{\prime }$ are
determined by (\ref{ronula}). Then $j_{-}^{\mu }$ in $S$ is obtained by the
LT, and it is given by (\ref{jotmi}). Hence, we find from (\ref{charge})
that
\begin{equation}
Q_{AB}=(1/c)\int_{0}^{l}(j_{+}^{0}+j_{-}^{0})dx=(1-\gamma )\rho _{0}l
\label{charAB}
\end{equation}%
This result is already found in [5]. Similarly, to find $Q_{AB}^{\prime }$
in $S_{AB}^{\prime }$ one first determines $j_{+}^{\mu }$ in $S$, $%
j_{+}^{\mu }=(c\rho _{0},0)$, and then by the LT one obtains $j_{+}^{\prime
\mu }=(c\gamma \rho _{0},-c\gamma \beta \rho _{0})$ in $S_{AB}^{\prime }$. $%
Q_{AB}^{\prime }$, for the moving loop in $S_{AB}^{\prime }$, is found from (%
\ref{charge}) in the same way as in [16] and it is, of course, equal to $%
Q_{AB}$. The evaluation of $Q_{EF}$ in the $\overline{EF}$ side, which is
parallel to the $\overline{AB}$ side, proceeds in the same way as for $%
Q_{AB} $. However the relative velocity of the $S_{EF}^{\prime }$ frame, the
rest frame for the electrons in the $\overline{EF}$ side, and the $S$ frame
is now $v^{\mu }=(\gamma c,-\gamma v)$. Hence, $j_{-}^{\mu }$ is now $%
j_{-}^{\mu }=(-c\gamma \rho _{0},c\gamma \beta \rho _{0})$ and
\begin{equation}
Q_{EF}=(1/c)\int_{l}^{0}(j_{+}^{0}+j_{-}^{0})dx=-Q_{AB}.  \label{charEF}
\end{equation}%
Thus we find that there are charges $Q_{AB}$ and $-Q_{AB}$ on the sides $%
\overline{AB}$ and $\overline{EF}$, respectively, in the $S$ frame in which
the loop is at rest. According to (\ref{charge}) $Q_{EF}$ is an invariant
charge, hence $Q_{EF}^{\prime }$ for the moving loop is $=Q_{EF}=-Q_{AB}$.
Moreover, it can be immediately concluded that the charge $Q_{BE}$, in $S$,
in the vertical side $\overline{BE}$must be the same as $Q_{AB}$; it is the
simple change of the spatial axes x and y. Thus $Q_{BE}=Q_{AB}$. Similarly
the charge $Q_{FA}$, in $S$, in the vertical side $\overline{FA}$ is $%
Q_{FA}=-Q_{AB}$. The total charge in $S$, the rest frame of the loop with
current, is zero $Q=Q_{AB}+Q_{BE}+Q_{EF}+Q_{FA}=0$, as it must be. It
remains zero in every IFR since the charges in all sides are invariant
charges according to (\ref{charge}), which means that they are the same for
both, moving and stationary current loop. Thus, we find the same behaviour
for a moving loop with current and for the same loop but at rest in a given
IFR.

In the traditional \textquotedblleft AT relativity\textquotedblright\
approach with the synchronous definition of length and the Lorentz
contraction (see, e.g., [21]) the charges on all sides are supposed to be
zero in the rest frame of a loop with current. In the IFR in which the loop
with current is moving it is found that $Q_{AB}^{\prime }\neq Q_{AB}$, due
to the Lorentz contraction, and that the charge on the $\overline{EF}$ side
is \textquotedblright -\textquotedblright\ of that one on the $\overline{AB}$
side. Furthermore, it is obtained that the charges on the vertical sides of
a loop with current are zero for both the loop at rest and moving loop,
since for vertical sides there is no Lorentz contraction. Such results
obtained in the common approach led the physics community (I am not aware of
any exception) to conclude that there is an electric moment $\mathbf{P}$ for
a moving loop with current, (see [21], Eq.(18-58)). The appearance of this
\textquotedblleft relativistic\textquotedblright\ effect and its
consequences are discussed in numerous papers and books. Contrary to all
these works we find in the \textquotedblleft TT
relativity\textquotedblright\ that \emph{at points far from that current
loop such a distribution of charges always (in any IFR) behaves like an
electric dipole, but as a 4D geometric quantity.} \newline
\newline
\newline
\textbf{5. CONCLUSIONS}\newline
\newline
In this paper the \textquotedblleft TT relativity\textquotedblright\ is
consistently applied to the investigation of relatively moving systems. We
presented the expressions for the covariantly defined spacetime length (\ref%
{covlen}) and for covariant 4D Lorentz transformations (\ref{fah}) when both
are written in geometrical terms, and in the \textquotedblleft
e\textquotedblright\ and \textquotedblleft r\textquotedblright\
coordinatizations. It is also shown that it is not the spatial length but
the spacetime length (\ref{covlen}) which is well defined quantity from the
\textquotedblleft TT relativity\textquotedblright\ viewpoint; in the rest
frame of the object and when the temporal part of $l_{AB}^{a}$ is zero the
spacetime length $l$ becomes the rest spatial length of the object. Using
this result and the covariant definition of charge (\ref{charge}) the
expressions (\ref{jotmi}) for the current density 4-vectors of a CCC were
found in the ions' rest frame. Then the 4-vectors $E^{\alpha }$ and $%
B^{\alpha }$ for a CCC are determined by means of the known $F^{\alpha \beta
}$ (\ref{elten}) and the relations (\ref{vezaEF}) and (\ref{vezFE}), which
connect the $F^{\alpha b}$ and $E^{\alpha },B^{\alpha }$ formulations of
electrodynamics. This yields Eq. (\ref{eovi}), which is one of the main
results found in this paper. It shows that for the obsevers at rest in the
ions' rest frame the spatial components $E^{i}$ of $E^{\alpha }$ are
different from zero outside a CCC. The second important result, which is
also found in a completely covariant manner, i.e., in the \textquotedblleft
TT relativity\textquotedblright\ treatment, is the existence of invariant
charges on a loop with current. There are opposite charges on opposite sides
of a square loop with current, but the total charge of that loop is zero.
These results are different from those found in all previous works in which
mainly the \textquotedblleft AT relativity\textquotedblright\ is used.%
\newline
\newline
\textbf{Acknowledgements. }I am indebted to Prof. F. Rohrlich for reading
the first version of this paper and for useful suggestions and
encouragement, and also to an anonymous referee for useful comments.\bigskip
\bigskip

\noindent \textbf{REFERENCES\bigskip }

1. T. Ivezi\'{c}, \textit{Found. Phys. Lett}. \textbf{12}, 105 (1999).

2. F. Rohrlich, \textit{Nuovo Cimento B} \textbf{45}, 76 (1966).

3. T Ivezi\'{c}, preprint SCAN-9802018, on the CERN server.

4. A. Gamba, \textit{Am. J. Phys}. \textbf{35}, 83 (1967).

5. T.Ivezi\'{c}, \textit{Phys. Lett. A} \textbf{144}, 427 (1990).

6. T.Ivezi\'{c}, preprint SCAN-9802017 (on the CERN server).

7. R.M. Wald, \textit{General relativity} (The University of Chicago Press,

Chicago, 1984).

8. C. Leubner, K. Aufinger and P. Krumm, \textit{Eur. J. Phys.} \textbf{13,}
170 (1992).

9. G. Cavalleri and C. Bernasconi, \textit{Nuovo Cimento B} \textbf{104},
545 (1989).

10. R. Anderson, I. Vetharaniam, G.E. Stedman, \textit{Phys. Rep}. \textbf{%
295},

93 (1998).

11. D.E. Fahnline, \textit{Am. J. Phys}. \textbf{50}, 818 (1982).

12. A. Einstein, \textit{Ann. Physik} \textbf{17, 891 }(1905), tr. by W.
Perrett

and G.B. Jeffery, in \textit{The principle of relativity} (Dover, New York).

13. K. Geiger, \textit{Phys. Rep.} \textbf{258,} 240 (1995).

14. M. Pardy, \textit{Phys. Rev. A} \textbf{55,} 1647 (1997).

15. A. Laub, T. Doderer, S.G. Lachenmann and R.P. Huebener,

\textit{Phys. Rev. Lett}. \textbf{75,} 1372 (1995).

16. N. Bili\'{c}, \textit{Phys. Lett. A} \textbf{162,} 87 (1992).

17. L. Baroni, E. Montanari and A.D. Pesci, \textit{Nuovo Cimento B} \textbf{%
109},

1275 (1994).

18. W.F. Edwards, C.S. Kenyon and D.K. Lemon, \textit{Phys. Rev. D} \textbf{%
14,}

922 (1976).

19. D.K. Lemon, W.F. Edwards and C.S. Kenyon, \textit{Phys. Lett. A }\textbf{%
62},

105 (1992).

20. T. Ivezi\'{c}, \textit{Phys. Rev. A} \textbf{44,} 2682 (1991).

21. W.K.H. Panofsky and M. Phillips, \textit{Classical electricity }

\textit{and magnetism, }2nd edn. (Addison-Wesley, Reading, Mass., 1962).

\end{document}